# A first-principles study on the effect of oxygen content on the structural and electronic properties of silicon suboxide as anode material for Lithium Ion Batteries

Obaidur Rahaman*,[1] Bohayra Mortazavi[1], and Timon Rabczuk[1,2]

[1]*Institute of Structural Mechanics, Bauhaus-Universität Weimar, Marienstr. 15, D-99423 Weimar, Germany*

[2]*School of Civil, Environmental and Architectural Engineering, Korea University,136-713 Seoul, Republic of Korea*

Abstract

Silicon suboxide is currently considered as a unique candidate for lithium ion batteries anode materials due to its considerable capacity. However, no adequate information exist about the role of oxygen content on its performance. To this aim, we used density functional theory to create silicon suboxide matrices of various Si:O ratios and investigated the role of oxygen content on the structural, dynamic, electronic properties and lithiation behavior of the matrices. Our study demonstrates that the O atoms interact strongly with the inserted Li atoms resulting in a disintegration of the host matrix. We found that higher concentration of oxygen atoms in the mixture reduces its relative expansion upon lithiation, which is a desirable quality for anode materials. It helps in preventing crack formation and pulverization due to large fluctuations in volume. Our study also demonstrate that a higher oxygen content increases the lithium storage capacity of the anode. However, it can also cause the formation of stable complexes like lithium silicates that might result into reversible capacity loss as indicated by the voltage-composition curves. The study provides valuable insights into the role of oxygen in moderating the interaction of lithium in silicon suboxide mixture in microscopic details.

*Corresponding author Email:  ramieor@gmail.com

Tel: +49 176 35365179

## 1. INTRODUCTION

Although graphite-based anode material is commonly used in commercial production of rechargeable Lithium Ion Batteries (LIB), it has several limitations. It is relatively expensive to manufacture and a specific capacity of 372 mAh/g imposes a limitation on the energy density. Silicon-based anode materials have been the recent focus of research in this area because of its abundance, low manufacturing cost and high theoretical lithium capacity of 4200 mAh/g.[1-3] Although it is recently emerging as a replacement for carbon-based anode materials, it also has some limitations. The biggest limitation of silicon-based anode material is its large volume change (>300%) during the lithiation and delithiation process leading to crack formation, pulverization and capacity fading as a consequence of loss of electrical contact.[4] Many experiments were conducted with altered design parameters or mixing silicon with other elements with an aim to overcome this problem.[5,6]

Mixing silicon with oxygen atoms is one of the proposed schemes to improve the performance of silicon-based anodes.[7,8] Kim et al. used inductively coupled plasma to investigate the effect of oxidation on the silicon oxide based anode and recommended the reduction of oxygen concentration below 18 atom % in order to increase the initial capacity.[9] Abel et al designed partially oxidized (≈13 atom % O) nano-structured silicon thin films that produced a high capacity of approximately 2200 mAh/g, which was reversibly cycled for 120 rounds with practically no capacity fade and 80% of the initial reversible capacity was retained after 300 cycles.[10]

Although several experimental studies were conducted in order to improve the quality of the silicon-suboxide anode material, an in-depth atomic-scale understanding of the interactions, bonding mechanism, energetic, diffusivity, mixture formations and their lithiation behavior is necessary for the rational design of silicon suboxide based anodes. Therefore, computational methods like *ab initio molecular dynamics* (AIMD) simulations are extensively used to design and investigate the performance of silicon suboxide since they can be very useful in discerning these microscopic properties.[11-13]

In a recent study, Chou et al. used Density Functional Theory (DFT) to investigate the lithiation behavior of silicon-rich oxide ($SiO_{1/3}$).[11] At high lithiation of $SiO_{1/3}$ matrices, they observed the formation of six-fold coordinated $[Li_6O]^{4+}$ clusters, as opposed to the irreversible formation of $Li_2O$ and various lithium silicates at a higher O:Si ratio ($SiO_2$ for example) leading to irreversible capacity loss. They suggested that the formation of $[Li_6O]^{4+}$ clusters could be related to the Si:O atomic ratio and the O spatial distribution in the suboxide mixture. They also suggested that the capacity and cyclability could be sensitive to the structural arrangement of the suboxide mixture. Thus, fine tuning the concentration and distribution of O atoms is crucial in maximizing the performance of the silicon suboxide based anode.

Following these suggestions we conducted a systematic study of the effect of O:Si ratio on the mixture formation and structural evolution with lithiation process. Various amorphous matrices of silicon suboxide with increasing Si:O ratio were generated. The geometrical, volumetric, dynamic and energetic properties were analyzed and compared in order to illustrate their sensitivities to the Si:O ratio. The factors important for designing a silicon suboxide anode material for optimum performance in LIB are discussed.

## 2. COMPUTATIONAL METHODS

Six amorphous silicon suboxide matrices $a$-SiO$_y$ (y=0, 0.1, 0.2, 0.3, 0.4 and 0.5) were created by six independent sequences of annealing, quenching and equilibration steps. Each system contained 30 Silicon atoms and the number of oxygen atoms varied according to the Si:O ratios. For each case, two independent supercells were used to improve the statistics. The atoms were randomly inserted in cubic simulation boxes followed by 4 ps of annealing at T=1500K using AIMD simulations. NVT ensembles with Langevin thermostat and a time step of 1 fs were used for the simulations. During the annealing process, the atoms were allowed to freely rearrange. After annealing, the systems were quenched at a rate of 0.3 K/fs to T=300K using NPT ensemble. This was followed by 4 ps of NPT simulation at T=300 K to allow for equilibration and 4 ps of NVT simulation for production.

The lithium intercalated amorphous silicon suboxide structures, $a$-Li$_x$SiO$_y$, were simulated by randomly adding lithium ions in the simulation box followed by 4 ps of NPT simulation for equilibration and 4 ps of NVT simulation for production. The lithium ions were added in four steps (x=1, 2, 3, 4) until saturation. The compositions of the $a$-Li$_x$SiO$_y$ structures are described in Table 1.

The DFT calculations were performed using *Vienna ab Initio Simulation Package*.[14,15] The Generalized Gradient Approximation (GGA) and Perdew-Wang 91 (PW91) functional were applied to represent the core electronic structure.[16] The interaction between valence and core electrons was described by the all electron frozen core, Projector Augmented Wave (PAW) method. All atoms were fully minimized with conjugate gradient method. We employed an energy cutoff of 400 eV and a k point mesh size of 2 × 2 × 2 in the Monkhorst-Pack scheme[17] for the Brilloin zone sampling, validated to be sufficient for the highly disordered $a$-Li$_x$SiO$_y$ systems in a previous study.[11]

## 3. RESULTS AND DISCUSSION

### 3.1 Analysis of the structure

Figure 1 shows the different amorphous structures of $Li_xSiO_y$ system for x=0, 1, 2, 3, 4 and y=0.5. It can be seen that the Li, Si and O atoms are well dispersed in the amorphous structures. A gradual disintegration of the initial $SiO_{0.5}$ structure (contacts between Si and O) is evident at increasingly higher level of Li atom insertion. Although many of the O atoms are surrounded by Li atoms, some are still interacting with Si atoms.

We note that the microscopic structure of silicon suboxide is complex and still under controversy. The experimental characterizations of silicon suboxide microstructures reveal that it can exist in both, homogenous and *a*-Si/*a*-SiO$_2$ nanocomposite forms.[10] The homogenous structure is stable under ambient conditions.[10] It is worthy to mention that *a*-Si/*a*-SiO$_2$ nanocomposite structures can be acquired through prolonged annealing of the initial homogenous form at high temperature which can lead to disproportionation of silicon suboxide into a mixture of *a*-Si and *a*-SiO$_2$.[18-22] Other computational studies were devoted in understanding the complex behavior of silicon suboxide in microscopic details, for instance, the formation of Si nanocrystals in Si suboxide composite was studied by a Monte Carlo method.[23] Due to the complex nature of the nanocomposite structures, it is almost impossible to build a representative volume element using only a few hundred atoms, a prerequisite for a DFT calculations. Consequently, in this study we only considered homogeneous structures of silicon suboxide by randomly placing Si and O atoms in a simulation box. To this aim, we build computational models using quick annealing and quenching steps, in which low contents of O atoms (x ranged from 0 to 0.5 in $SiO_x$) were randomly dispersed in the simulation box forming

Si-O-Si units and creating a homogenous mixture of these two atoms. Chou et al. observed such random distributions of O atoms in their DFT study of $SiO_{1/3}$.[11] Other first principles studies used similar methods to construct the initial configurations of silicon suboxide based materials.[11-13] This is also in agreement with the experimental evidence of homogenous incorporation of oxygen in nanostructured silicon thin films under similar conditions.[10]

In addition, we note that we constructed the supercell with 30 atoms for the smallest system (*a*-Si). However, its size at full lithiation (x=4) is four times bigger and it becomes prohibitively expensive for an *ab initio* molecular dynamics simulation calculation. A large number of calculations with multiple steps and samples are necessary for obtaining a reliable statistics on the properties of the materials under study. Considering these limitations we adopted the smallest supercell of 30 atoms which is in the ballpark of other studies on similar systems.[11,24,25] The validation of using such a system for atomistic simulations can be found elsewhere.[25]

In order to gain further insight into the atomic interactions, we calculated the pair distribution functions (PDFs) of all the amorphous structures. Figure 2 shows the Si-Si, Si-O, Si-Li and O-Li PDFs of the $Li_xSiO_y$ systems at complete saturation, i.e. x=4. The broad peaks are indicative of the amorphous nature of the matrices. A strong O-Li interaction, even at low oxygen concentration, is evident from the dominant peaks around 1.9 Å. A peak for Si-O is only visible at higher concentration of O. The peak is at 1.78 Å for y=0.3, slowly shifting to 1.68 Å for y=0.5. The peak is increasingly sharper at higher concentration of O. A comparison of O-Li and Si-O peaks indicates that although O-Li interaction is prominent in all cases, the Si-O interaction becomes comparable at higher O concentration. The O concentration has no significant effect on the first peaks of Si-Si and Si-Li at 2.45 Å and 2.63 Å, respectively.

Figure 3 shows the coordination numbers (CN) of the Si-Si, Si-O, Si-Li and O-Li atom pairs in the $a$-Li$_x$SiO$_{0.5}$ matrices at cutoff values of 3.0 Å, 2.0 Å, 3.5 Å and 2.75 Å respectively. Each cutoff value approximately corresponds to the position of the first minimum after the first peak in the radial distribution function. It can be clearly seen that both the Si-Si and Si-O CNs gradually decrease with increasing concentration of lithium atoms in the mixture. The Si-Si CN falls from 3.77 to 1.01 and the Si-O CN drops from 1.04 to 0.2. On the other hand the Si-Li CN gradually increases from 6.12 at x=1 to a final value of 11.05 at x=4. The O-Li CN also increases from 3.55 at x=1 to 6.09 at x=4. These results indicate that the oxygen atoms prefer to interact with the lithium atoms rather than the Si atoms, specially at high Li concentration. The insertion of Li gradually disintegrates the $a$-SiO$_{0.5}$ host mixture with O atoms coordinating with 6 Li atoms at saturation. These results are in agreement with the previous DFT study on $a$-SiO$_{1/3}$.[11] The evolution patterns of the CNs are similar in other matrices studied in this work (data not shown).

Angular distributions can be useful in describing the local geometrical structures of the matrices. The Si-O-Si, Si-O-Li and Li-O-Li angular distributions were calculated and their evolution with increasing concentration of Li were investigated. No dramatic changes in the angular distributions were found with matrices of different O concentrations. However, systematic changes in the angular distributions were observed with the insertions of Li atoms. To illustrate the general trends the angular distributions were shown for two cases: LiSiO$_{0.5}$ and Li$_4$SiO$_{0.5}$ (Figure 4). Si-O-Si distribution has a sharp peak at 123° and a minor peak at 99° for LiSiO$_{0.5}$. No Si-O-Si species were observed for Li$_4$SiO$_{0.5}$ (although they were still observed for Li$_2$SiO$_{0.5}$ and Li$_3$SiO$_{0.5}$ matrices, data not shown). This is in accordance with the observation of the previous section that the insertion of Li atoms drastically disintegrates the host mixture of SiO$_y$, specially at the saturation point. For LiSiO$_{0.5}$, the Si-O-Li angles are widely distributed with a peak around 90° and a shoulder around 130°. However, for Li$_4$SiO$_{0.5}$, the peak is sharper and there is a second peak at 163° instead of a shoulder. This indicates an increase of ordering in the mixture with increasing Li concentration. The Li-O-Li distribution has a major peak around 80° and a minor peak around 145° for both LiSiO$_{0.5}$ and Li$_4$SiO$_{0.5}$ without a significant variation between them (the distribution looks similar also for Li$_2$SiO$_{0.5}$ and Li$_3$SiO$_{0.5}$ matrices, data not shown). The results put together suggest that the saturation by lithium insertion causes the disintegration of Si-O mixture replaced by a strong ordered Li-O interaction.

## 3.2 Evolution of volume and density

A large volume expansion of the electrode material is related to poor capacity retention of the battery. In addition large alterations in the volume causes cracking and pulverization. Thus, the volume expansion due to the insertion of Li atoms is a parameter of prime importance for the application of SiO$_y$ mixture as an negative-electrode material. In general a low relative volume expansion upon lithiation is desirable over a high relative volume expansion. Figure 5a) shows the fractional changes in the volume (V- V$_0$)/V$_0$, where V$_0$ and V are the volume of the mixture before and after the insertion of Li respectively. For the case of *a*-Si (without any O), the increase in volume upon lithiation matches well with previous experimental[26] and computational studies.[24] At complete saturation (x=4) the fractional change in volume reached up to 3.14 times its initial volume. This is rather a large change in volume. However, as it can be seen from the figure, the presence of O in the mixture reduces this volume expansion. In fact the higher the O content the lower is the volume expansion. This trend probably indicates a tighter packing of O and Li as compared to the packing of Si and Li in the amorphous matrices. Thus, it can be inferred that a higher O content is desirable in terms of volume expansivity.

Next we analyzed the evolution of density with increasing Li concentration. The density of the mixture gradually decreases with the increasing amount of Li atoms in the host mixture (Figure 5b)). The decrease in density upon lithiation gradually reduces with increasing O content. The density of the mixture with highest O concentration ($Li_4SiO_{0.5}$) has the highest density at saturation (x=4). Since O atoms are lighter than Si atoms, an increment in the O content in the mixture should reflect in a lower density as long as both types of atoms occupy the same volume. However, we rather see an increase in density with higher O content, particularly in the presence of Li (x=1,2,3 and 4) than its absence (x=0) in the mixture. Thus, it can be inferred that the higher density of the mixture with higher O content indicates a tighter packing of the atoms induced by the O, specially in the presence of Li. This suggests a strong O-Li interaction in agreement with previous observations.

**3.3 Diffusion**

Next we look at the diffusion of different elements (Si, O and Li) in the $Li_xSiO_{0.5}$ systems. Figure 6 shows the Root Mean-Square Deviation (RMSD) of different elements in the last two picoseconds of the NVT production step at room temperature. Before the insertion of Li, the RMSD of both Si and O were around 0.2 Å. At higher concentrations of Li the RMSD increased a bit, to about 0.3. It is interesting to note that the RMSD of Si and O are comparable in each stage of Li insertion. Although the insertion of lighter and smaller Li atoms slightly increased the diffusivity of Si and O, it did not increase further with higher concentrations of Li. At each stage of lithiation, the RMSD of Li was distinguishably higher than those of Si and O, as expected. However, it can also be seen that the RMSD of Li slightly decreased (from 0.56 to 0.44) with increasing concentration of Li. One possible explanation of this could be the filling of the empty Si-Si and Si-O interspatial regions by the small Li atoms, increasingly lowering the allowed spaces for Li atoms to diffuse to. No general trends in the RMSD could be discerned for different concentrations of O atoms in different $a$-Li$_x$SiO$_y$ systems.

**3.4 Density of State**

In order to investigate the effect of O on the electronic properties of the mixture, the Density of States (DOS) for Si, SiO$_{0.5}$, Li$_4$Si and Li$_4$SiO$_{0.5}$ were compared (figure 7). The Fermi energy was shifted to zero of the energy scale and shown using a dotted vertical line. A small band gap is clearly visible for the case of Si which characterizes its semiconducting behavior. The band gap is significantly shortened in the case of SiO$_{0.5}$. Additional electronic states were also introduced in the conduction band. At the saturation level of Li concentration both Li$_4$Si and Li$_4$SiO$_{0.5}$ matrices show metallic states as demonstrated by a lack of band gap. This is in agreement with previous studies that suggest a drastic shift of Si from semiconductor state to metallic state upon insertion of even small amount of Li.[27,28] On the other hand, the presence of the O atoms in Li$_4$SiO$_{0.5}$ created a large gap in the valence band which is absent in Li$_4$Si. No noteworthy general trends were observed with increasing concentration of O atoms in the matrices. Thus, the effect of insertion of Li and O in the $a$-Si mixture can be summarized as follows: 1) addition of Li transforms the mixture from a semiconductor to a metallic state 2) addition of O reduces the available electronic states in the valence band and introduces new electronic states in the conduction band.

**3.5 Formation energy**

In order to further explore the effect of O content on the electronic properties of the matrices we calculated the formation energies and lithiation voltages. The formation energies of the $a$-Li$_x$SiO$_y$ systems can be approximated by the following equation:

$$E_f = E_{Li_xSiO_y} - (xE_{Li} + E_{SiO_y}) \quad (1)$$

assuming that the free energies of these systems can be approximated by their total energies at 0K, neglecting the entropy and pressure terms. The terms $E_{Li_xSiO_y}$ and $E_{SiO_y}$ correspond to the energies of $a$-Li$_x$SiO$_y$ and $a$-SiO$_y$ systems per Si atom. $E_{Li}$ is the energy of a body-centered-cubic Li (bcc-Li) system per Li atom. The later quantity was estimated using a supercell of 64 Li atoms with a 4 × 4 × 4 in the Monkhorst-Pack scheme starting from the crystalline bcc structure. The formation energies of the $a$-Li$_x$SiO$_y$ systems are shown in figure 8. The formation energies of all the amorphous matrices decrease monotonically with increasing concentration of Li atoms until saturation (x=4). The formation energies of the matrices containing O atoms are relatively lower than that of the matrices without any O atoms (Li$_x$Si systems, y=0). It can also be seen that the descending trends of the formation energy curves are increasingly steeper with increasing concentration of O. This indicate that the incorporation of O atoms in the amorphous Si makes the Li incorporation more energetically favorable (as reflected by a more negative formation energy). This can be explained by the fact that the Li-O interaction is stronger than the Li-Si interaction as suggested in this work. This trend is in agreement with the previous DFT investigation where the formation energy curve for $a$-Li$_x$SiO$_{1/3}$ was observed to be steeper than that of $a$-Li$_x$Si.[11]

**3.6 Voltage**

A third order polynomial was fitted to each of the formation energy curves. The potentials of the $a$-Li$_x$SiO$_y$ systems were derived using the following equation applied on the polynomial fitting:

$$V = -\frac{dE_f(x)}{dx} \qquad (2)$$

Figure 9 shows the voltage-composition curves of the $a$-Li$_x$SiO$_y$ systems. The lithiation voltage of the $a$-Si (y=0) is predicted to be within 0.2 and 0.6 V which is in good agreement with the previous DFT results[29] and well within the range of values obtained by an experimental study on sputtered $a$-Si.[30] The lithiation voltage, gradually increases with increasing concentration of O atoms, significantly at early stage of lithiation (up to 1.26 V for $a$-SiO$_{0.5}$) and moderately (up to 0.4 V for $a$-SiO$_{0.5}$) at later stage of lithiation at x=4. This is consistent with our earlier observation that the presence of O increases the favorable incorporation of Li. This is also in agreement with the DFT study that suggested a higher lithiation voltage for $a$-SiO$_{1/3}$ in comparison to $a$-Si, especially at the early stage of lithiation.[11] This is also in agreement with an experimental study that suggested a higher lithium insertion potential with increasing oxygen concentration in silicon suboxide.[10] The lithiation voltage at and around half way to saturation (x=2) falls almost within the experimental range for $a$-Si, even for the cases of relatively high O concentration (y=0.5).

The lithiation voltages for all cases are still within the desirable range of LIB negative electrode application except the ones at higher oxygen concentration. For instance at y=0.3, 0.4 and 0.5, a rise in the lithiation voltage during the charging cycle, although desirable, might be associated with the formation of irreversible species of Li$_2$O and lithium silicates. The formation of such irreversible species can cause entrapment of Li and consequently a loss of reversible capacity. Further studies are necessary to investigate the formation of such species and the possible role of O on their formation mechanisms.

## 4. CONCLUSIONS

In this work, we used density functional theory to study the microscopic properties of $Li_xSiO_y$ systems to understand of the effect of O content on the performance of silicon suboxide as an negative-electrode material for LIB. Analyses of the Radial Distribution Functions and coordination numbers revealed that the Li atoms prefer to form bonds with O atoms instead of Si atoms, in agreement with a previous study on $SiO_{1/3}$.[11] The analysis of volume expansion clearly indicated a tighter packing of the mixture with higher concentration of O. This is a desirable quality of a potential negative-electrode material for LIB because of its lower risk of pulverization and cracking due to the excessive alteration of volume. The analysis of formation energy indicates an increasingly favorable lithium incorporation with increasing O content, another desirable property of negative-electrode material. However, very high concentration of O atom in the mixture generates high lithiation voltage that might be associated with the formation of stable complexes causing irreversible capacity loss. Within the framework of this study an optimum concentration of O for the best performance of silicon suboxide as a negative-electrode material in LIB was not detected. However, this study characterizes the role of O content on several properties of silicon suboxide that are crucial in determining its performance as a negative-electrode material. We believe that the knowledge and insights obtained in this work can be useful toward the rational design of a silicon suboxide based anode material with optimum concentration of oxygen.

**Acknowledgements**: The authors gratefully acknowledge the financial support of the European Research Council (Grant number 615132).**References**

(1) Zhang, Y.; Li, Y.; Wang, Z.; Zhao, K. *Nano Letters* **2014**, *14*, 7161.
(2) Nguyen, C. C.; Choi, H.; Song, S. *Journal of the Electrochemical Society* **2013**, *160*, A906.

**Figures**

Figure 1

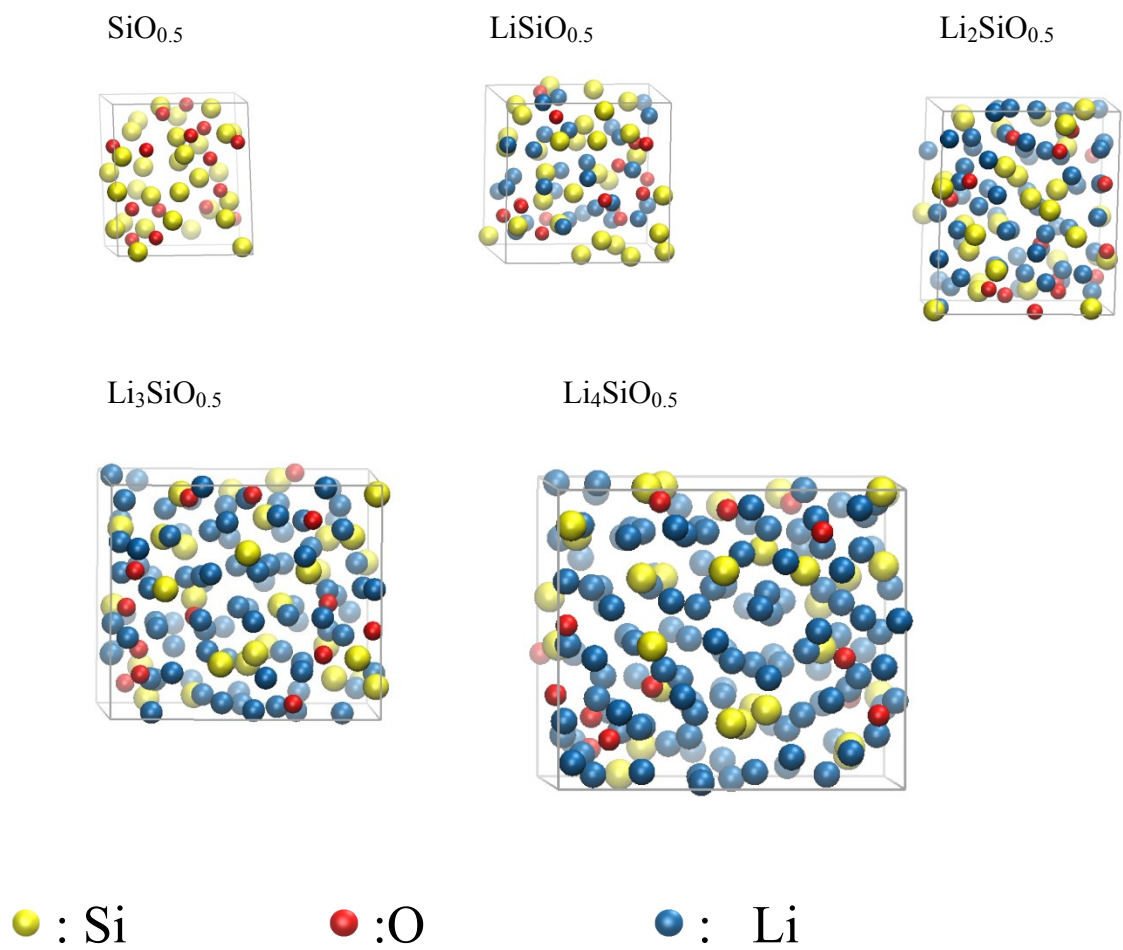

Figure 2

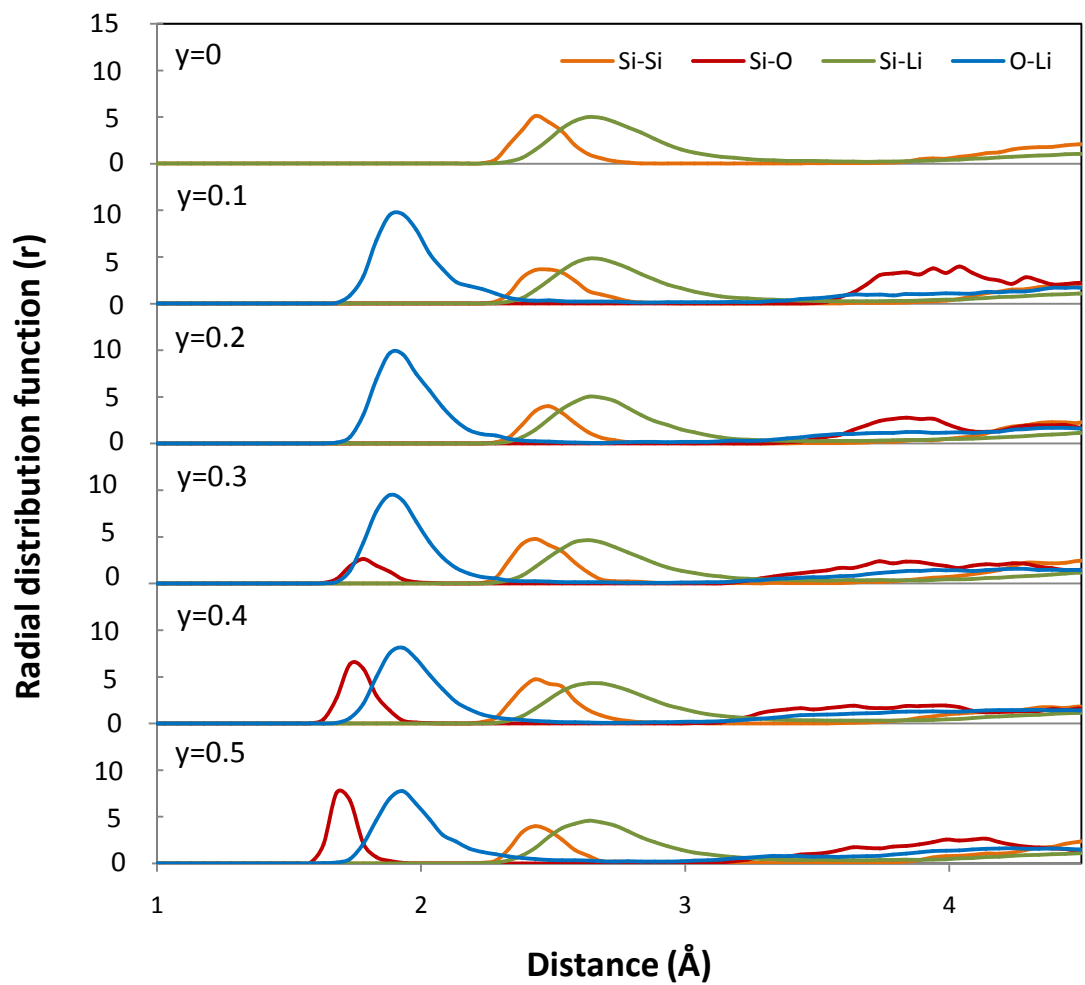

Figure 3

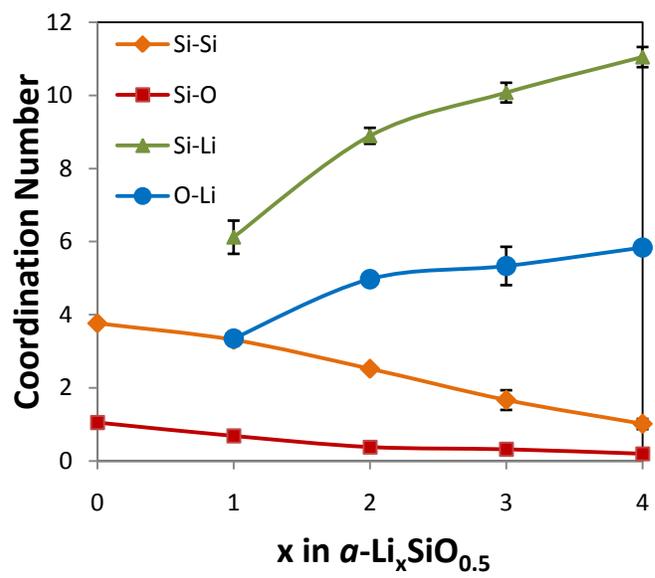

Figure 4

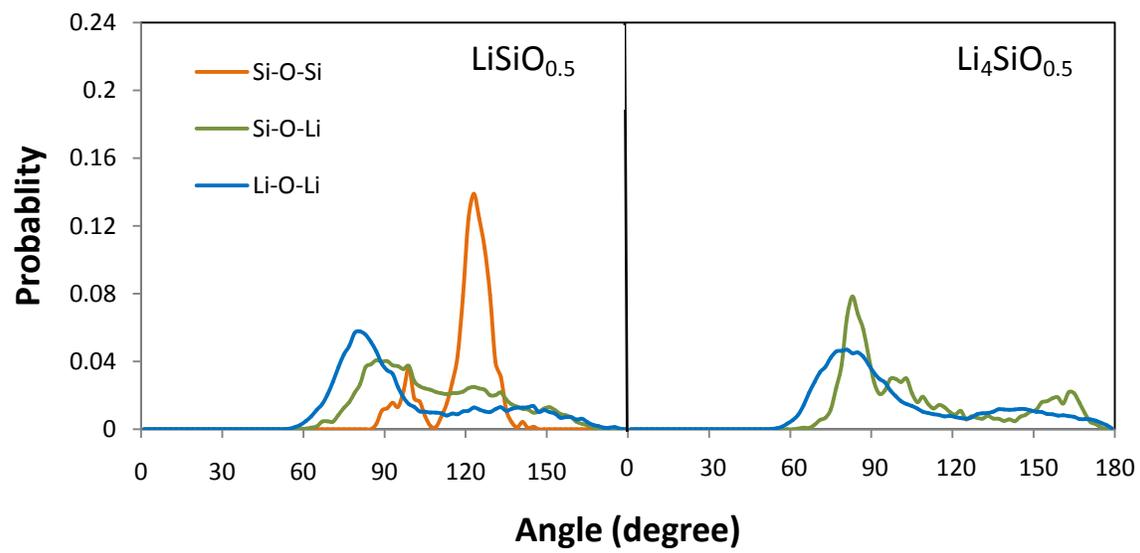

Figure 5

a)

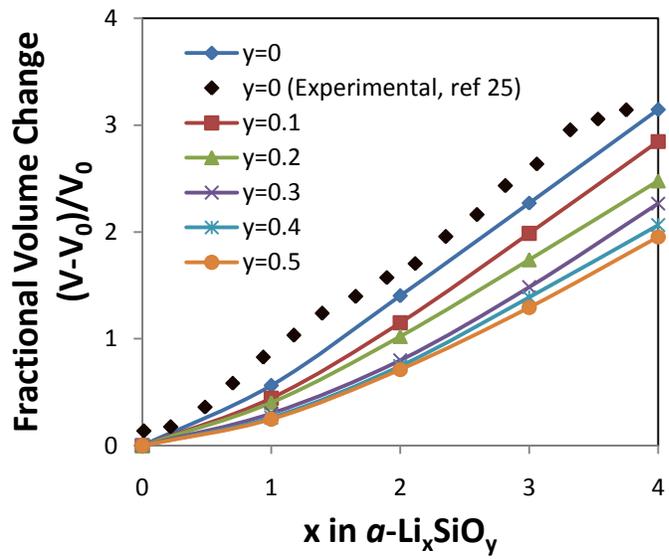

b)

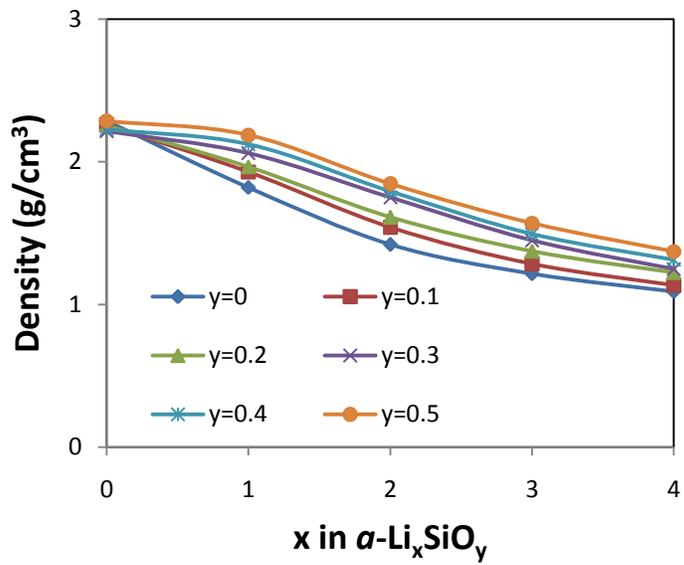

Figure 6

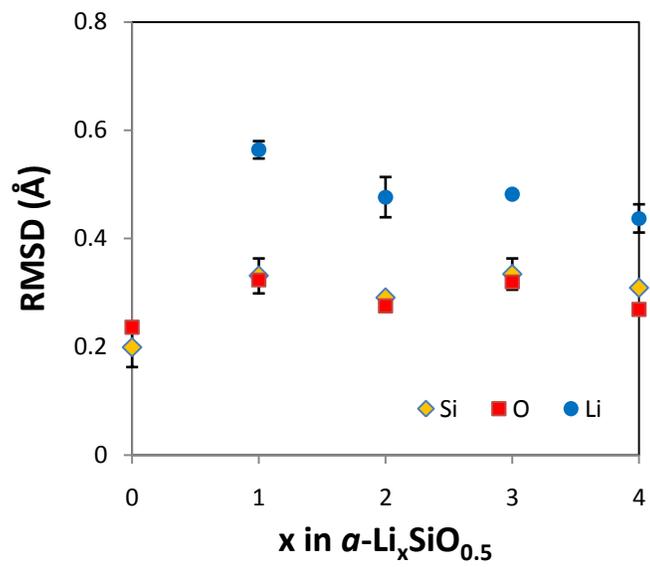

Figure 7

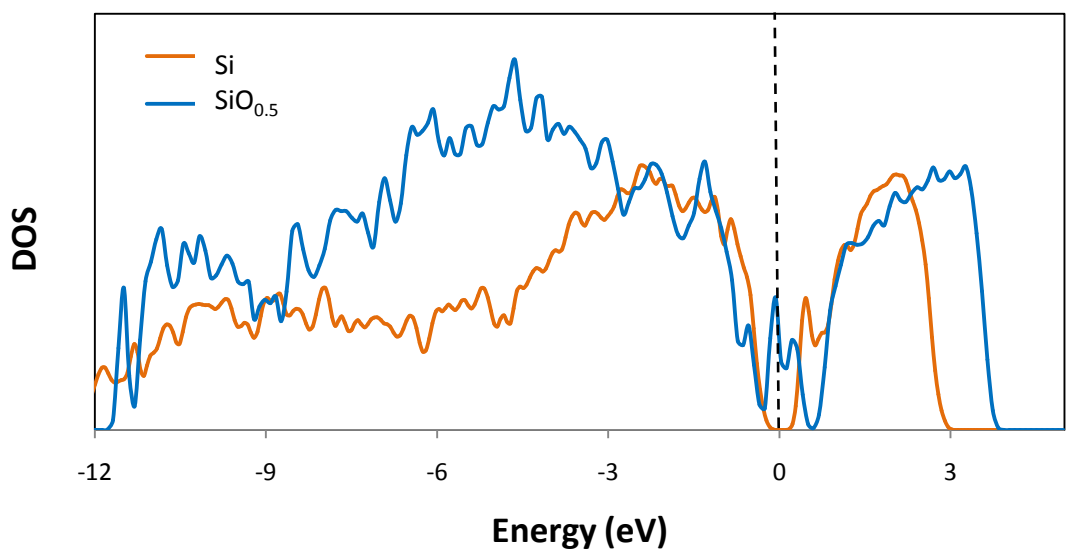

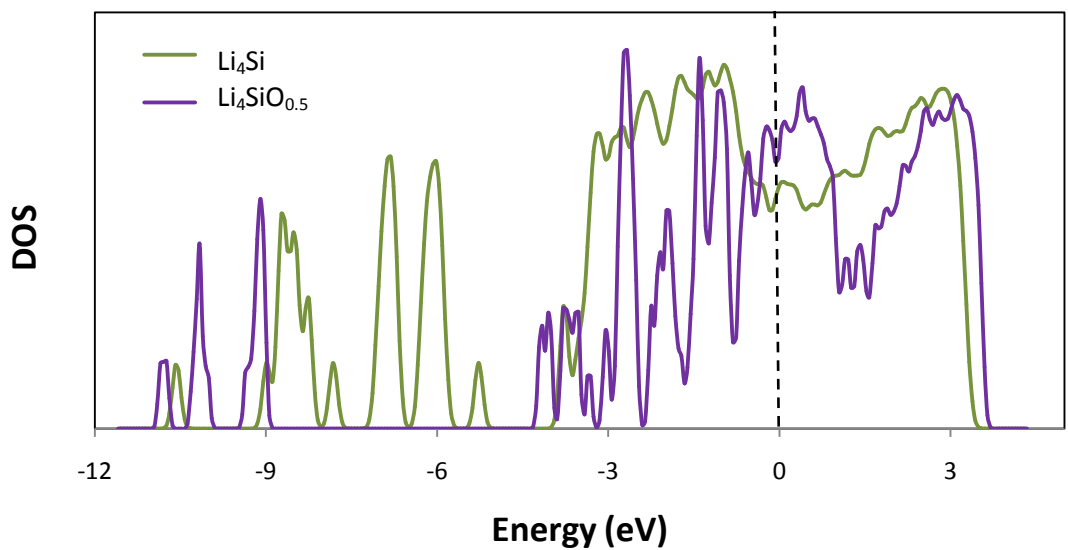

Figure 8

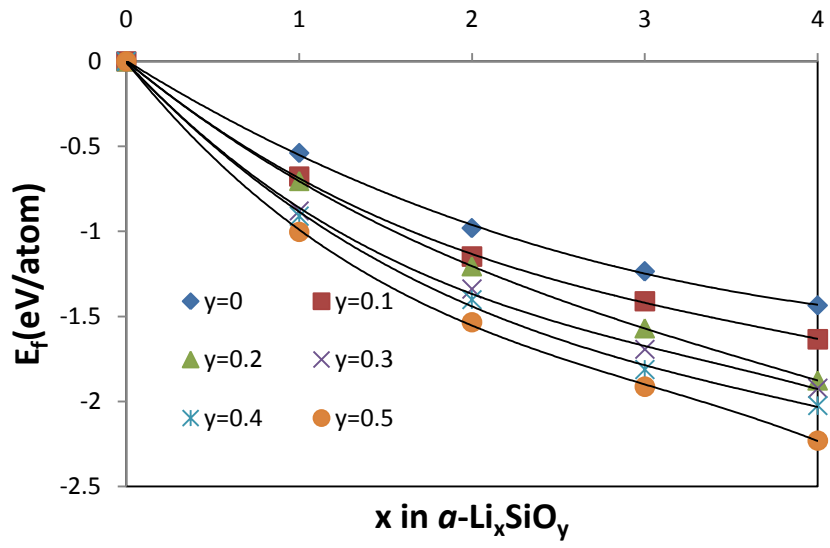

Figure 9

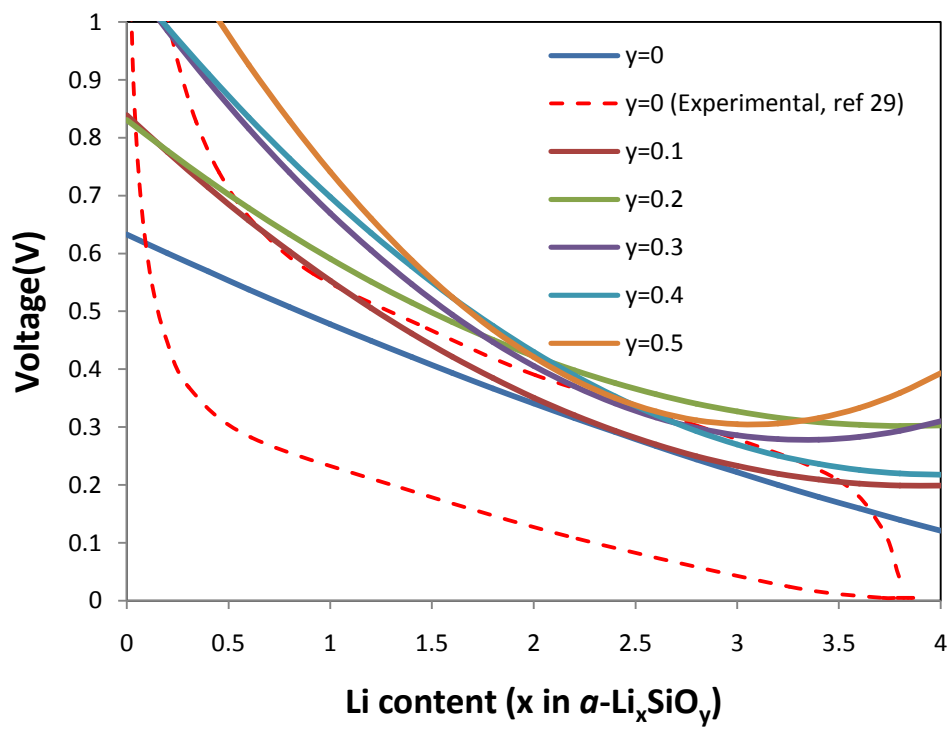